\documentclass[12pt]{article}
\usepackage{epsf}
\usepackage{amsmath}
\usepackage{graphics}
\usepackage{cite}

\renewcommand{\thefootnote}{\#\arabic{footnote}}
\begin{document}

\newcommand{\gtrsim}{ \mathop{}_{\textstyle \sim}^{\textstyle >} }
\newcommand{\lesssim}{ \mathop{}_{\textstyle \sim}^{\textstyle <} }

\newcommand{\rem}[1]{{\bf #1}}

\renewcommand{\thefootnote}{\fnsymbol{footnote}}
\setcounter{footnote}{0}
\begin{titlepage}

\def\thefootnote{\fnsymbol{footnote}}

\begin{center}
\hfill January 2015\\
\vskip .5in
\bigskip
\bigskip
{\Large \bf Cyclic Entropy: An Alternative to Inflationary Cosmology}

\vskip .45in

{\bf Paul Howard Frampton}

{\em Courtyard Hotel, Whalley Avenue, New Haven, CT 06511, USA.}

\end{center}

\vskip .4in
\begin{abstract}
We address how to construct an infinitely cyclic universe model.
A major consideration is to make the entropy cyclic which requires
the entropy to be re-set to zero in each cycle expansion $\rightarrow$ turnaround
$\rightarrow$ contraction $\rightarrow$ bounce $\rightarrow$ etc. 
Here we re-set entropy at the turnaround by selecting
the introverse (visible universe) from the extroverse which is generated by the
accelerated expansion. In the model, the observed homogeneity is
explained by the low entropy at the bounce, The observed flatness
arises from the contraction together with the reduction in size
between the expanding and contracting universe. The
present flatness is predicted to be very precise.
\end{abstract}
\end{titlepage}

\renewcommand{\thepage}{\arabic{page}}
\setcounter{page}{1}
\renewcommand{\thefootnote}{\#\arabic{footnote}}

\newpage

\section{Introduction}

\noindent
It is of broad interest among all physicists, indeed all scientists, to understand 
the early universe especially whether a singular bang ever occurred as
suggested by the most simple-minded interpretation of the
Friedmann equation. The idea of a big bang has proved remarkably
resilient despite the knowledge that extrapolating back the Friedmann
equation to the singularity must somehow be flawed.

\bigskip

\noindent
The discovery\cite{Penzias} of the cosmic microwave background(CMB) 
resolved a dichotomy then existing in theoretical cosmology between 
steady-state and big-bang theories. The interpretation of the CMB as a relic of a big bang was
compelling and the steady-state theory died. Actually in 1965 it was
really a trichotomy being reduced to a dichotomy because a third theory,
a bounce in a cyclic cosmology, had been under study since 1922
\cite{Friedmann}. 

\bigskip

\noindent
An infinitely cyclic cosmology was always the most popular alternative to
an initial singularity but its correct development was hindered by
an impossible-seeming stumbling block created by the monotonically
increasing entropy
of the universe. Entropy is sometimes perceived as not a
fundamental concept because it involves a very large number
of particles and is therefore not a property of a single particle.
Nevertheless, in studying theoretical cosmology we shall argue
that the entropy of the universe is the single most important
concept. Interest in the pursuit of a successful cyclic model
is reflected in the existence of three recent popular books
\cite{SteinhardtBook,FramptonBook,PenroseBook}, as well
as in \cite{KCWali}.

\bigskip

\noindent 
Nevertheless, for purely theoretical reasons, the cyclic model 
had been discarded due to the Tolman Entropy Conundrum(TEC)
\cite{Tolman1,Tolman2}. The TEC, stated too simply, is that the entropy
of the universe necessarily increases, due to the second law of
thermodynamics, and therefore cycles become larger and longer in the
future, smaller and shorter in the past, implying that a big bang must
have occurred at a finite time in the past.

\bigskip

\noindent
Some progress towards a solution of the TEC was made in \cite{BF}
using the Come Back Empty (CBE) assumption in the so-called BF model.
A huge entropy was there jettisoned at turnaround and the significantly
smaller universe, empty of matter, contracted adiabatically to a bounce
with zero entropy. The original BF model employed so-called phantom dark energy
with equation of state $\omega < -1$. Since this violates every well-known 
energy condition,
it appears more sensible to employ a cosmological constant with $\omega = -1$
as we insist upon here, in a standard $\Lambda CDM$ model of the present
era . We shall find that CBE at turnaround is still possible. An abbreviated 
presentation of some of the present results appeared in \cite{Frampton}.

\bigskip

\noindent
The emphasis in this article is on entropy and the second law of thermodynamics.
No cause or mechanism will be provided for the turnaround from expansion to
contraction of the bounce from contraction to expansion. The only theory
used will be general relativity, in particular the Friedmann equation to describe
the expansion and contraction era of the space-time described by a
homogeneous and isotropic FRLW metric. At almost all times general
relativity is sufficient. For the turnaround and bounce, an unknown extra
ingredient is necessary as a part of a C-theory where C means Complete.
Our assumption is that use of the Friedmann equation is sufficient to
draw conclusions about the cyclic cosmology.

\bigskip

\noindent
To illustrate a phenomenological fudge factor to precipitate the turnaround and
bounce one could modify the Friedmann equation to
\begin{equation}
\left( \frac{\dot{a(t)}}{a(t)} \right)^2 = \frac{8 \pi G}{3} \rho_{TOT}(t)
\left[ \left(1 - \frac{a(t)^2}{a(t_T)^2} \right) \left(1 - \frac{a(t_B)^2}{a(t)^2} \right) \right]
- \frac{k(t)}{a(t)^2}
\label{fudge}
\end{equation}
where $t_T$, $t_B$ are the turnaround, bounce times respectively and
where the fudge factor in square brackets has no derivation and violates
diffeomorphism invariance so is merely to illustrate a point, in the absence
of any known C-theory. In \cite{CBECCC} it is shown that infinite cyclicity
requires that $t_T \sim 1.3 $ Ty, at which time the radius of the extroverse
is $R_{EV}(t_T) \sim 10^{67}$ m, a length scale far beyond where
general relativity has been tested. At the bounce, if it is at the Planck time,
the radius of the extroverse is $R_{EV}(t_B=10^{-44}s) \sim 10^{-6}$ m,
which is a length scale where Newton's law of gravity has not even been
tested.
 
\bigskip

\noindent
Eq.(\ref{fudge}) is a good approximation to the Friedmann equation
\begin{equation}
\left( \frac{\dot{a(t)}}{a(t)} \right)^2 = \frac{8 \pi G}{3} \rho_{TOT}(t)
- \frac{k(t)}{a(t)^2}
\label{Fried}
\end{equation}
for all times except very close to $t \sim t_T$ or $t \sim t_B$ because $a(t_T)$
is extremely large and $a(t_B)$ extremely small. In this paper we shall use only
Eq.(\ref{Fried}) and make the reasonable assumption 
that corrections from effects of C-theory near to
the turnaround and bounce do not significantly change conclusions about
entropy and the second law of thermodynamics.

\bigskip

\noindent
Most of research on cosmology and the early universe in the last few decades
have assumed an inflationary paradigm where there is a short burst of
superluminal accelerated expansion early on in the universe. The present
cyclic model does not use any such assumption. 
An inflationary stage is redundant because the geometry becomes flat
without it. This is because the value $\Omega_{TOTAL}(t)=1$ is a stable
fixed point of the contraction equation, just as it is an unstable fixed point
under expansion. This fact alone is sufficient to achieve the observed 
flatness without inflation. The CBE model makes the flatness much more exact to a degree
that the departure from $\Omega_{TOT}(t_0)=1$ at the present time
is too small to be observable.

\bigskip

\noindent
The plan of this paper is that in Section 2 we discuss the present expansion era;
in Section 3 we study the turnaround from expansion to contraction; in Section 4
the topic is the bounce from contraction to expansion; finally, in Section 5,
there is a concluding discussion.

\newpage

\section{Present expansion era}

\noindent
Assuming the cosmological principle of homogeneity and isotropy leads
to the FLRW metric
\begin{equation}
ds^2 = dt^2 - a(t)^2 \left[ \frac{dr^2}{1-k(t) r^2} + r^2 (d\theta^2 + \sin^2\theta d\phi^2) \right],
\end{equation}
where $a(t)$ is the scale factor and $k$ the curvature. Inserting this metric
into the Einstein equation leads to two Friedmann equations. The first is the expansion
equation
\begin{equation}
H(t)^2 = \left(\frac{\dot{a}(t)}{a(t)} \right)^2 = \frac{8\pi G}{3} \rho_{TOT}(t) - \frac{k(t)}{a(t)^2},
\label{Friedmann1}
\end{equation}
where $\rho_{TOT}(t)$ is the total density.

\bigskip

\noindent
Using the continuity equation $\dot{\rho}(t) + 3H(t)(\rho(t)+p(t))=0$, differentiation
of Eq.(\ref{Friedmann1}) gives rise to a second equation
\begin{equation}
\frac{\ddot{a}(t)}{a(t)} = - \frac{4\pi G}{3} (\rho(t) + 3p(t) ).
\label{Friedmann2}
\end{equation}

\bigskip

\noindent
The critical density is defined by $\rho_c(t) = [3H(t)^2/8\pi G]$ and discussion
of flatness involves the proximity to unity of the quantity
\begin{equation}
\Omega_{TOT} (t) = \frac{\rho_{TOT}}{\rho_c(t)}
\end{equation}

\bigskip

\noindent
Let the present time be $t=t_0\simeq1.38\times10^{10}$ and normalize $a(t)$ by $a(t_0)=1$. Other useful cosmic times, all measured relative to the
would-have-been bang at $t=0$, are the Planck time at $t_{Planck}=10^{-44}$s,
the electroweak phase transition time at $t_{EW}=10^{-10}$s, the onset of matter domination
at $t_m=4.7\times10^4y$ and the onset of dark energy domination at $t_{DE}=9.8$Gy.

\bigskip

\noindent
In Table 1 we summarize the well-established history of the present expansion era.

\bigskip

\newpage

\begin{table}[htdp]
\caption{History of present expansion.}
\begin{center}
\begin{tabular}{|c|c|c|}
\hline
TIME  & Scale factor & Comments    \\
t & a(t) & \\
\hline
\hline
13.8Gy          &   1                     & present time           \\
\hline
$9.8Gy < t < 13.8Gy$         &    $a(t) =\exp[H_0(t-t_0)] $            & dark energy dominated         \\
\hline
$t_{DE}=9.8Gy$  & 0.75 & onset of dark energy domination  \\
\hline
 $47ky < t < 9.8Gy$ & $a(t)\propto t^{2/3}$ & matter dominated \\
\hline
$47ky$ & $2.1 \times 10^{-4}$ & onset of matter domination\\
\hline
$t < 47ky$  & $a(t) \propto t^{1/2}$ & radiation dominated \\
\hline
$t=10^{-10}s$ & $1.7\times 10^{-15}$ & electroweak phase transition\\
\hline
$10^{-44}s < t < 10^{-10}s$ &$a(t)\propto t^{1/2}$ & possible inflation or bounce\\
\hline 
$t = 10^{-44}s$ & $1.7\times 10^{-32}$ & Planck time \\
\hline
\hline
\end{tabular}
\end{center}
\label{expansiontable}
\end{table}

\noindent
In the unadorned big bang theory, one has dependences of the scale factor $a(t) \sim t^{1/2}$
for $t_{Planck}<t<t_m$, $a(t)\sim t^{2/3}$ for $4t_m<t<t_{DE}$, and
$a(t) = \exp[H_0(t-t_0)]$ for $t_{DE}<t$. This leads to the values of $a(t)$
displayed in Table 1.

\bigskip

\noindent
The two most striking observations of the visible universe are isotropy to an
accuracy $1\pm O(10^{-5})$ and \cite{Dicke} flatness $\Omega_{TOT}(t_0)=1.00\pm0.05$.
In big bang theory the latter implies that in the early universe $\Omega_{TOT}(t)$ is given 
at $t=t_{EW},t_{Planck}$ by
\begin{equation}
\left| \Omega_{TOT}(t_{EW})  - 1\right| \sim O(10^{-26}),
\label{flat1}
\end{equation}
\begin{equation}
\left| \Omega_{TOT}(t_{Planck}) = 1\right| \sim O(10^{-60}),
\label{flat2}
\end{equation}
as we shall re-derive later by considering the contraction of a cyclic
universe. Once
we incorporate cyclic entropy, the resultant flatness will be even more
extreme at the Planck time than shown in Eq.(\ref{flat2}), see {\it e.g.}
Eq.(\ref{OmegaPL3}) below for an example of this phenomenon.

\bigskip

\noindent
We shall focus first on the geometrical condition of flatness, then return
towards the end to address homogeneity in terms of, what is an intimately 
related concept, the low entropy of the universe at the bounce.

\bigskip

\noindent
A convenient way to discuss flatness is to rearrange Eq.(\ref{Friedmann1})
after division by $H(t)^2$ as
\begin{equation}
(\Omega_{TOT}(t) - 1) = \frac{k(t)}{\dot{a}(t)^2}.
\label{Flatness}
\end{equation}

\bigskip

\noindent
In a decelerating expansion the denominator of the RHS in
Eq.(\ref{Flatness}) becomes smaller and smaller and $\Omega_{TOT}(t)$ deviates
more and more from $\Omega_{TOT}(t)=1$. This is why the proximity
of $[\Omega_{TOT}(t_0)-1]$ to zero imposes the strong initial conditions in
Eqs.(\ref{flat1}) and (\ref{flat2}). 

\bigskip

\noindent
The most popular explanation of flatness is ``inflation" 
\cite{Starobinsky,Guth1,Linde,AS,LL}
which can be defined as insertion of a brief period 
of highly accelerated superluminal expansion
at a time $t \simeq t_{inflation}$ during the radiation-dominated era
$t_{Planck}<t_{inflation}<t_{EW}$. While inflating, $\dot{a}(t)$ becomes extremely
large enforcing flatness sufficiently so precisely in Eq.(\ref{Flatness}) 
that the subsequent decelerating
expansion during $t_{inflation}<t<t_{DE}$ does not remove it.

\bigskip

\noindent 
Inflation also explains homogeneity and has other successes
including the prediction of 
scale-invariant density perturbations. 
It has been claimed that inflation has been successfully
incorporated into string theory
\cite{KKLT,KKLMMT}. The successful discovery of the BEH boson adds
credibility to the existence of the scalar inflaton. One possible objection to
inflation, that it leads to eternal inflation\cite{Guth2} and hence to
a multiverse in which predictivity is hampered by the measure
problem\cite{Vilenkin}, is not a fatal flaw.
Only a compelling alternative theory could cast doubt on the 
correctness of inflation: cyclic entropy can provide
such an alternative as discussed in the remainder of
this paper.

\bigskip

\noindent
If one insisted on criticizing inflation (any theory can be criticized!), it
would be that the initial conditions before inflation are unspecified. 

\bigskip

\noindent

\newpage

\section{Turnaround from expansion to contraction}

\noindent
To solve the TEC, the entropy of the introverse immediately after turnaround must
essentially vanish as discussed in \cite{BF}. As an aid to discussion of cyclic entropy, 
we display Tables 2 and 3 which illustrate the future evolution of $\Lambda CDM$ models. 

\bigskip

\noindent
In Table 2 we start from the present time $t=t_0=13.8Gy$ and proceed into
the future first in steps of $10Gy$, then $50Gy$, then $250Gy$ and finally $500Gy$.
In the second column is the corresponding value of the scale factor $a(t)$
with the conventional normalization $a(t_0) = 1$.

\bigskip

\noindent
In the third column, we assume the number of galaxies inside
the introverse at present, $t=t_0$, is exactly one trillion, $10^{12}$. Any departure would
involve a corresponding rescaling of all the entries in this column, but
the entries $\leq1$ will be unchanged by any reasonable such departure. 

\bigskip

\noindent 
For the fourth column in Table 2, the radius
of the introverse (IV) is calculated as usual from
\begin{equation}
R_{IV}(t) = 44Gly + c \int_{t_0}^{t_T} dt ~~  a(t)^{-1},
\label{Visible}
\end{equation}
with $a(t)= \exp[H_0(t-t_0)]$. The value of $R_{IV}(t)$ gradually evolves
to an asymptotic value of $R_{IV}(\infty)=58Gly$, limited by the speed of light.

\bigskip

\noindent
Because of the superluminal expansion
of space, $R_{IV}(t_0)$ is increased to the much larger radius
$R_{EV}(t)=a(t)R_{IV}(t_0)$ by the stretching of space, as shown in the fifth columnn
of Table 2. We refer to this larger space as the extroverse (EV).

\bigskip

\noindent
It now requires very great care to identify the appropriate introverse at
turnaround to fulfill \cite{BF} the CBE assumption necessary for cyclic entropy. 
One might initially
be tempted to include
the Milky Way galaxy which we inhabit. That would be too Ptolemaic and 
a most grievous error! 
The correct choice of
introverse at $t=t_T$ is instead a sphere which contains no matter at all, luminous 
or dark, including no black holes. It contains instead only dark 
energy with no entropy and 
small quantities of both curvature and radiation, this last being
inevitable and actually crucial to the ensuing derivation of flatness.

\bigskip

\noindent
In the discussions of \cite{BF}, a central role was played by the fraction ($f$) 
defined by
\begin{equation}
f=\frac{R_{IV}(t_T)}{a(t_T)R_{EV}(t_0)} = \frac{R_{IV}(t_T)}{R_{EV}(t_T)}
\label{f}
\end{equation}
This all-important quantity is listed in the sixth and last column of Table 2.

\bigskip
\bigskip

\noindent
Because the present time $t=t_0$ is not special, we have taken the liberty
to display Table 3 which is a slightly different but more physical version of Table 2.
Table 3 begins at the physically important
time $t=t_{DE}=9.8Gy$ which is the onset of dark energy
domination and hence of the superluminal accelerated expansion.
We highlight the second row which is the present time $t=t_0=t_{DE}+4Gy$,
because it illustrates that a extroverse presently exists with $R_{EV}=52Gly$
and volume $\sim 60\%$ larger than the introverse. This must
contains some
600 billion additional galaxies which will remain forever unobservable, given a
continued accelerated expansion.

\bigskip

\begin{table}[htdp]
\caption{Dependence on $t_T$ normalized to $R_{IV}(t_0)=R_{EV}(t_0)$.}
\begin{center}
\begin{tabular}{||c|c|c|c|c|c||}
\hline
Turnaround & $a(t_T)$ & Galaxies & $R_{IV}(t_T)$ & $R_{EV}(t_T)$ & f \\
($t_T-t_0$) & &in introverse   & &  &     \\
\hline
\hline
0          &    $1$ &$10^{12}$                    & $44Gly$& $44Gly$ &  1   \\
\hline
10Gy         &  $2.1$               &  $1.8\times10^{11}$    & $51Gly$   & $91Gly$  & $0.56$ \\
\hline
20Gy & $4.3$ &$2.4\times10^{10}$ & $55Gly$ & $190Gly$  & $0.29$ \\
\hline
30Gy & $8.8$ & $3.0\times10^9$ & $56Gly$ & $390Gly$ & $0.14$\\
\hline
40Gy & $18$ & $3.6\times10^8$ & $57Gly$ & $800Gly$ & $0.071$ \\
\hline
50Gy & $37$ & $4.8\times10^7$  & $58Gly$ & $1.6Tly$ & $0.036$ \\
\hline
100Gy & $ 1.5\times10^3$ & $710$ & $58Gly$ & $66Tly$ & $8.8\times10^{-4}$ \\
\hline
150Gy & $5.7\times10^4$ & $\leq1$ & $58Gly$ & $ 2.3Ply$  & $2.5\times10^{-5}$ \\
\hline 
200Gy & $2.0\times10^6$ & $\leq1$ & $58Gly$ & $87 Ply$ & $6.7\times10^{-7}$ \\
\hline
250Gy & $7.4\times10^7$ & $\leq1$ & $58Gly$ & $3.2Ely$ & $1.8\times10^{-8}$ \\
\hline
500Gy  & $5.4\times10^{15}$ & $\leq1$ & $58Gly$ & $240Yly$ & $2.4\times10^{-16}$ \\
\hline
1Ty  &  $3.0\times 10^{31}$ & $\leq1$ & $58Gly$ & $1.3\times10^{33}Gly$ & $4.5\times10^{-32}$  \\
\hline
\hline
\end{tabular}
\end{center}
\label{tTTable}
\end{table}

\begin{table}[htdp]
\caption{Dependence on $t_T$ normalized to $R_{IV}(t_{DE})=R_{EV}(t_{DE})$.}
\begin{center}
\begin{tabular}{||c|c|c|c|c|c||}
\hline
Turnaround & $a(t_T)$ & Galaxies & $R_{IV}(t_T)$ & $R_{EV}(t_T)$ & f \\
($t_T-t_{DE}$) & &in introverse  & &  &     \\
\hline
\hline
0          &    $0.75$ &$1.6\times10^{12}$        & $39Gly$& $39Gly$ &  1   \\
\hline
\hline
4Gy  &  1  & $10^{12}$  & $44Gly$ & $52Gly$    & $0.85$ \\
\hline
\hline
10Gy         &  $1.6$               &  $5.5\times10^{11}$    & $49Gly$   & $70Gly$  & $0.70$ \\
\hline
20Gy & $3.2$ &$8.8\times10^{10}$ & $53Gly$ & $141Gly$  & $0.38$ \\
\hline
30Gy & $6.6$ & $1.1\times10^{10}$ & $56Gly$ & $290Gly$ & $0.19$\\
\hline
40Gy & $13$ & $1.6\times10^9$ & $57Gly$ & $570Gly$ & $0.10$ \\
\hline
50Gy & $28$ & $1.4\times10^8$  & $57Gly$ & $1.2Tly$ & $0.047$ \\
\hline
100Gy & $ 1.1\times10^3$ & $2800$ & $58Gly$ & $48Tly$ & $1.2\times10^{-3}$ \\
\hline
150Gy & $4.3\times10^4$ & $\leq1$ & $58Gly$ & $ 1.9Ply$  & $3.1\times10^{-5}$ \\
\hline 
200Gy & $1.5\times10^6$ & $\leq1$ & $58Gly$ & $66Ply$ & $8.8\times10^{-7}$ \\
\hline
250Gy & $5.5\times10^7$ & $\leq1$ & $58Gly$ & $2.4Ely$ & $2.4\times10^{-8}$ \\
\hline
500Gy  & $4.0\times10^{15}$ & $\leq1$ & $58Gly$ & $180Yly$ & $3.2\times10^{-16}$ \\
\hline
1Ty & $2.2\times10^{31}$ & $\leq1$ & $58Gly$ & $9.7\times10^{32}Gly$ & $6.1\times10^{-32}$ \\
\hline
\hline
\end{tabular}
\end{center}
\label{tTTable2}
\end{table}

\bigskip
\bigskip

\newpage

\section{Bounce from Contraction to Expansion}

\noindent
The CBE contracting universe contains no matter.
It does, however, contain a known amount of radiation and in
approaching the bounce this radiation dominates the energy and entropy. 
During the expansion era, at $t=t_0$ its energy contribution is
$\Omega_{\gamma}(t_0) = 1.3 \times 10^{-4}$. Its entropy will be discussed below.

\bigskip

\noindent
The normal Friedmann equation without cyclic entropy is, from Eq.(\ref{Friedmann2}) and using 
the radiation equation of state
$3p=\rho$
\begin{equation}
\frac{\ddot{a}(t)}{a(t)} = - \frac{8\pi G}{3} \rho_{\gamma}(t),
\label{contraction}
\end{equation}
and we need to calculate $\dot{a}(t_B)$ which appears in this contraction
version of Eq.(\ref{Flatness})
\begin{equation}
\left| \Omega_{TOT}(t_B) - 1 \right| = \left| \frac{k}{\dot{a}(t_B)^2}\right|.
\label{Omega1}
\end{equation}

\bigskip

\noindent
Using 
\begin{equation}
\rho_{\gamma}(t) = \frac{\rho_{\gamma}(t_0)}{{a}(t)^4}
\label{radiation}
\end{equation}
in Eq.(\ref{contraction}) leads to the following
\begin{equation}
\frac{d\dot{a}(t)}{dt} = -\frac{8\pi G \rho_{\gamma}(t_0)}{3} a(t)^{-3}.
\label{radeq}
\end{equation}

\bigskip

\noindent
During the radiation-dominated era we have, from Table 1
that
\begin{equation}
a(t) = C_{\gamma} t^{1/2},
\label{rad}
\end{equation}
where
\begin{equation}
C_{\gamma} = 2.3 \times 10^{-10}  s^{-1/2}.
\label{radcoeff}
\end{equation}

\bigskip

\noindent
Now we can use
\begin{equation}
\Omega_{\gamma}(t_0) = \frac{8\pi G\rho_{\gamma}(t_0)}{3 H(t_0)^2}
\label{Omegagamma}
\end{equation}

\noindent
and 
\begin{equation}
|k| = H(t_0)^2
\label{k}
\end{equation} 

\noindent 
to rewrite Eq.(\ref{radeq}) in the more useful form
\begin{equation}
\int d(\dot{a}) = \frac{C_{\gamma}^6}{4 H(t_0)^4 \Omega_{\gamma}(t_0)^2} \int dt ~~t^{-3/2} 
\label{integral}
\end{equation}
and inserting
\begin{equation}
|k| = H(t_0)^2
\label{k}
\end{equation}
in Eq.(\ref{Omega1}) and performing the integrals gives the flatness result
\begin{equation}
\left| \Omega_{TOTAL}(t_B) -1 \right| = C_{\Omega} t_B,
\label{Omega2}
\end{equation}

\bigskip

\noindent
in which the coefficient, $C_{\Omega}$, is readily calculated to be
\begin{equation}
C_{\Omega} = \frac{C_{\gamma}^6}{4 H(t_0)^2 \Omega_{\gamma}(t_0)^2} = 3.9\times 10^{-16} ~~s^{-1}.
\label{Omegacoefficient}
\end{equation}

\bigskip

\noindent
Eqs.(\ref{Omega2}) and (\ref{Omegacoefficient}) now give, 
ignoring prefactors which are $O(1)$, at the electroweak
and Planck times, $t_{EW} = 10^{-10}s$ and $t_{Planck} = 10^{-44}s$, respectively

\begin{equation}
\left| \Omega_{TOTAL}(t_{EW}) - 1 \right| ~~ \sim ~~  O(10^{-26})
\label{OmegaEW}
\end{equation}
\noindent
and
\begin{equation}
\left| \Omega_{TOTAL}(t_{Planck}) - 1 \right| ~~ \sim ~~ O(10^{-60})
\label{OmegaEW}
\end{equation}

\noindent
which are nothing more than Eqs. (\ref{flat1}) and (\ref{flat2}) respectively, 
now rederived using time reversal acting on the expansion era.

\bigskip

\noindent
But the requirement of cyclic entropy importantly
adds one more multiplicative factor
on the right-hand-side of Eq.(\ref{Omega2}) because the CBE contracting universe is described
by a modified scale factor $\hat{a}(t) = f a(t)$, where f is the usually small fraction
given in the last column of Tables 2 and 3 and in the fourth column of
Table 4.

\bigskip

\noindent
With cyclic entropy, for the contracting universe, Eq.(\ref{radeq}) must
therefore be replaced by
\begin{equation}
\frac{d\dot{\hat{a}}(t)}{dt} = -\frac{8\pi G \rho_{\gamma}(t_0)}{3} \hat{a}(t)^{-3}.
\label{radeqhat}
\end{equation}

\bigskip

\noindent
Inserting now the defining CBE relationship $\hat{a}(t)=fa(t)$ we easily find that Eq.(\ref{Omega2}) inherits
a multiplicative factor which is a fourth power of the CBE contraction fraction, $f^4$, giving the new flatness result

\begin{equation}
\left| \Omega_{TOTAL}(t_B) -  1 \right| = f^4C_{\Omega} t_B.
\label{Omega3}
\end{equation}

\bigskip

\noindent
Let us illustrate the remarkable implications of Eq.(\ref{Omega3}) by two 
examples, using the results displayed in Tables 2 and 3.

\bigskip

\noindent
As a first example, consider the turnaround time $t_T$ satisfying
$(t_T-t_{DE})=150Gly$. This is the shortest time after which the visible
universe contains $\leq1$ galaxies and so is where the CBE assumption
first becomes straightforward to implement. For this case $f= 3.1\times 10^{-5}$
and hence, ignoring prefactors which are O(1), $f^4 \sim 10^{-18}$ and
therefore
\begin{equation}
\left| \Omega_{TOTAL}(t_{EW}) - 1 \right| ~~ \sim ~~  O(10^{-44})
\label{OmegaEW2}
\end{equation}
and
\begin{equation}
\left| \Omega_{TOTAL}(t_{Planck}) - 1 \right| ~~ \sim ~~ O(10^{-78}).
\label{OmegaPL2}
\end{equation}

\bigskip

\noindent
As a second example, consider the turnaround time $t_T$ satisfying
$(t_T-t_{DE}) = 1Ty$ so that, from Table 3, $f^4\sim 10^{-126}$.
This leads naturally to the remarkable initial conditions
\begin{equation}
\left| \Omega_{TOTAL}(t_{EW}) - 1 \right| ~~ \sim ~~  O(10^{-152})
\label{OmegaEW3}
\end{equation}
\noindent
and
\begin{equation}
\left| \Omega_{TOTAL}(t_{Planck}) - 1 \right| ~~ \sim ~~ O(10^{-186}).
\label{OmegaPL3}
\end{equation}

\bigskip

\noindent
The total energy density at the present time $t=t_0$ becomes extremely close to the critical
density in these two examples. It is respectively
\begin{equation}
\left| \Omega_{TOTAL}(t_0) - 1 \right| ~~ \sim ~~  O(10^{-18}),
\label{OmegaNow1}
\end{equation}

\noindent
for the choice $(t_T-t_{DE})=150Gy$,and
\begin{equation}
\left| \Omega_{TOTAL}(t_0) - 1 \right| ~~ \sim ~~ O(10^{-126}),
\label{OmegaNow2}
\end{equation}
for the case $(t_T-t_{DE})=1Ty$.

\bigskip

\noindent
From a practical and observational point of view, however, the density perturbations
of the CMB render it impossible to check Eqs.(\ref{OmegaNow1}) or (\ref{OmegaNow2}) to
more than five decimal places.

\bigskip
\bigskip

\noindent
Homogeneity at the bounce is ensured by low entropy.
Let us examine the value of the dimensionless entropy $S_{TOTAL}(t_B)/k$ which
originates only from electromagnetic radiation, $S_{TOTAL}(t_B)=S_{\gamma}(t_B)$.

\bigskip

\noindent
First note that at the present time the CMB in the visible universe has dimensionless entropy
given in Weinberg's book as \cite{Weinberg}
\begin{equation}
S_{\gamma}(t_0)/k = \left( \frac{2 \pi^2}{45}\right) g_* V_{VU}T^3 \sim 10^{88}.
\label{entropyCMB}
\end{equation}

\bigskip

\noindent
The radiation entropy remains constant during adiabatic contraction 
so the bounce entropy can be reliably estimated
by using the extensive property that entropy is proportional to volume:
\begin{equation}
S_{\gamma}(t_B)/k \sim 10^{88} f(t_T)^3,
\label{entropy}
\end{equation}
where we have displayed explicitly the important dependence of $f$ on the turnaround time $t_T$.

\bigskip

\noindent
Using Table 4, the bounce entropy for $(t_T-t_{DE}) = 150Gy$ is thus $S_{\gamma}(t_B)/k \sim 10^{74}$
which, while a very large number, is nevertheless
extremely small compared to the present 
total entropy of the visible universe, $S_{TOTAL}(t_0) \sim 10^{124}$, and may
be sufficiently small to ensure the observed homogeneity
and isotropy. 

\bigskip

\noindent
On the other hand, if $S_{\gamma}(t_B)/k \sim 10^{74}$ is still
insufficiently low, then as can be confirmed from the final column of Table 4, 
one may increase to a trillion years the turnaround time, $(t_T-t_{DE}) = 1Ty$, 
and thus render the bounce entropy identically zero, $S_{\gamma}(t_B)/k  = 0$,
whereupon homogeneity and isotropy can become exact.
 
\begin{table}[htdp]
\caption{Dependence of bounce entropy $S_{\gamma}(t_B)/k$ on $(t_T-t_{DE})$}
\begin{center}
\begin{tabular}{||c|c|c|c|c||}
\hline
Turnaround & $a(t_T)$ & Galaxies  & f & Bounce \\
($t_T-t_{DE}$) & &in introverse  & &  entropy  ($S_{\gamma}/k$)  \\
\hline
\hline
0          &    $0.75$ &$1.6\times10^{12}$      &  1 &  - \\
\hline
\hline
4Gy  &  1  & $10^{12}$    & $0.85$ &  -  \\
\hline
\hline
10Gy         &  $1.6$               &  $5.5\times10^{11}$     & $0.70$ & - \\
\hline
20Gy & $3.2$ &$8.8\times10^{10}$  & $0.38$ & - \\
\hline
30Gy & $6.6$ & $1.1\times10^{10}$ & $0.19$ & - \\
\hline
40Gy & $13$ & $1.6\times10^9$ & $0.10$ & - \\
\hline
50Gy & $28$ & $1.4\times10^8$  & $0.047$ & - \\
\hline
100Gy & $ 1.1\times10^3$ & $2800$  & $1.2\times10^{-3}$ & - \\
\hline
150Gy & $4.3\times10^4$ & $\leq1$ & $3.1\times10^{-5}$ & $\sim10^{74}$ \\
\hline 
200Gy & $1.5\times10^6$ & $\leq1$& $8.8\times10^{-7}$ & $\sim10^{70}$ \\
\hline
250Gy & $5.5\times10^7$ & $\leq1$ & $2.4\times10^{-8}$ & $\sim10^{65}$ \\
\hline
500Gy  & $4.0\times10^{15}$ & $\leq1$  & $3.2\times10^{-16}$ & $\sim10^{50}$ \\
\hline
1Ty & $2.2\times10^{31}$ & $\leq1$ & $6.1\times10^{-32}$ & $\sim0$ \\
\hline
\hline
\end{tabular}
\end{center}
\label{STable}
\end{table}

\newpage

\section{Discussion}

\noindent
Previous work on cyclic cosmology, especially back in the twentieth century,
was stymied by ignorance of the accelerated expansion which was
first discovered only in 1998. The no-go theorem or Tolman Entropy
Conundrum (TEC) implicitly assumed decelerated expansion and hence
no extroverse surrounding the introverse into which to jettison
the huge entropy which accumulates from irreversible processes 
during expansion.

\bigskip

\noindent
One prediction of the cyclic entropy model is that the presently
observed flatness which can be expressed as $|\Omega(t_0)-1|\simeq 0$ 
is precise, possibly being 
accurate to 18 or even 128 decimal places depending on the turnaround time
as exhibited in our Eqs.(\ref{OmegaNow1}) and (\ref{OmegaNow2}).
This precision arises from the Come Back Empty (CBE) assumption
\cite{BF} where the visible universe after turnaround generally has a radius only 
a very small fraction
$f$ of the radius of the previously expanding visible universe. Such theoretical
precision far exceeds that of any possible observational measurement of $\Omega(t_0)$.

\bigskip

\noindent
In summary, cyclic entropy provides an alternative to 
inflationary cosmology, at least as far as providing an alternative
explanation of the observed homogeneity and flatness. To be a more complete
alternative would require an explanation of the density perturbations
which seed structure formation. In inflationary cosmology these
are explained by quantum fluctuations of an inflaton field.
In the alternative of cyclic entropy, such fluctuations might 
reasonably be expected
to arise due to quantum effects during the contraction era.

\newpage

\section*{Acknowledgement}
Thanks are due to the University of Miami's Department of Physics
for kind hospitality while writing up this paper.

\end{document}